# Mitigating Denial of Service Attacks in Fog-Based Wireless Sensor Networks Using Machine Learning Techniques

Ademola P. Abidoye
aabidoye1@clayton.edu
Department of Computer Science
Clayton State University
Atlanta, USA

Ibidun C. Obagbuwa
ibidun.obagbuwa@spu.ac.za
Department of Computer Science and Information Technology
Sol Plaatje University
Kimberley, South Africa

Nureni A. Azeez
nazeez@unilag.edu.ng
Department of Computer Science
University of Lagos
Lagos, Nigeria

**Abstract**

Wireless sensor networks are considered to be among the most significant and innovative technologies in the 21st century due to their wide range of industrial applications. Sensor nodes in these networks are susceptible to a variety of assaults due to their special qualities and method of deployment. In WSNs, denial of service attacks are common attacks in sensor networks. It is difficult to design a detection and prevention system that would effectively reduce the impact of these attacks on WSNs. In order to identify assaults on WSNs, this study suggests using two machine learning models: decision trees and XGBoost. The WSNs dataset was the subject of extensive tests to identify denial of service attacks. The experimental findings demonstrate that the XGBoost model, when applied to the entire dataset, has a higher true positive rate (98.3%) than the Decision tree approach (97.3%) and a lower false positive rate (1.7%) than the Decision tree technique (2.7%). Like this, with selected dataset assaults, the XGBoost approach has a higher true positive rate (99.01%) than the Decision tree technique (97.50%) and a lower false positive rate (0.99%) than the Decision tree technique (2.50%).

## 1 Introduction

The rapid advancement of digital electronics, wireless communications, and micro-electro-mechanical systems has made it possible to create inexpensive, small, and affordable sensor nodes. A network of dispersed autonomous electronic devices that can be used to monitor environmental variables is known as a wireless sensor network (WSN) consisting of small nodes [1]. The small nodes can exchange information and communicate with one another. Typically, sensor nodes are placed to gather environmental data in the targeted locations such as sun radiation, ambient temperature, relative humidity, wind speed, pressure, and contaminants. Typically, this data is sent to data centers for additional processing, and the results are utilized to guide choices concerning the sensed region.

Recent advancements in camera sensors have prompted the creation of the first-person form of network known as visual sensor networks (VSNs). Without the advancements in wireless communication technology, VSNs have expanded WSN applications in ways we could not have predicted.

Environmental monitoring, medical applications, industrial applications, automotive sectors, smart parking systems, and other fields are some of the application areas of WSNs [2]. Introduction of fifth-generation (5G) mobile communications, WSNs are now of enormous relevance. This innovative technology enables greater machine-to-machine (M2M) communication inside the Internet of Things (IoT) environment, producing massive volumes of data every minute. To better understand environmental data, this data must be processed and analyzed using contemporary technologies like high-performance computing (HPC) and cloud computing [3]. Sensor nodes are a type of ad hoc network that may form multi-hop wireless networks and interact with one another without the need for pre-existing infrastructures or a centralized organization [4]. In an ideal world, such networks would require identification between nodes before transmitting sensed data to a data-gathering center. A typical sensor node is powered by small batteries. This has remained one of the most difficult obstacles in the adoption of this technology, particularly in places requiring a long lifetime of network use and a superior quality of service (QoS).

Many models and solutions for reducing sensor node energy consumption have been developed in the literature [5-7]. One way to minimize energy consumption in WSNs is data aggregation, which reduces the volume of data carried to a destination. In a multi-hop network, sensed data is pre-processed at intermediate or aggregator nodes. They combine the given data using aggregation functions [8] such as minimize, sum, and average, then send the results to the upper layer (cloud) for additional processing and storage. In contrast, data aggregation strategies may reduce QoS metrics in WSNs such as latency, data accuracy, and fault tolerance. Furthermore, data aggregation among sensor nodes makes sensed data more vulnerable to various threats such as denial of service assaults. A hacked node, for example, might either broadcast arbitrary data as aggregated data or reveal data obtained from other nodes in an unauthorized manner. Secure data transmission via a wireless medium necessitates an assessment of data confidentiality and integrity [9] to preserve transferred data and ensure that uncompromised data is routinely sent to the cloud for processing and storage. This is accomplished through i) data confidentiality, which ensures that only the intended destination node receives sensed data, ii) data integrity, which ensures that the content of sensed data is maintained during transmission between a sender node and a receiver node, and iii) data authentication, which ensures that the data's source is the intended sender node [10]. Every sensor network needs to design an effective security method to maintain dependable, trustworthy performance, preserve sensed data, and assure network component authenticity. WSN communication is through a multi-hop. An intruder exploits the weaknesses of the nodes to launch one or more of the attacks. Denial of service (DoS) attacks are common risks to WSN security. The purpose of these attacks is to make the network inaccessible to traditional sensor nodes. DoS attacks disrupt WSN functionality by flooding the target nodes with data packets. These attacks can be harmful if the nodes are employed in sensitive industries like environmental monitoring, military operations, and remote surveillance.

DoS attacks can damage any layer of a sensor network [11]. DoS attacks at the network layer, for example, can disable a network service, transmit compromised data, and cause network congestion by flooding the network with worthless packets. Similarly, an attacker may corrupt a sensor node's settings, rendering its resources inaccessible.

Sensor nodes in WSNs typically communicate via wireless medium to sense or monitor the environment, making them vulnerable to various foreign agents during deployment and opening the door for adversaries to launch attacks against the sensor network, either to destroy the nodes or capture the nodes to extract useful data.

An adversary can betray the confidentiality of the data received by the adversary nodes in standard point-to-point encryption methods where sensing data is encoded at each relaying node. The consequences of such attacks are severe: DoS assaults on such networks may result in real-world harm to people's health and safety if the network is used in the healthcare sector, as well as counterattacks in military operations. As a result, various security measures must be implemented to defend WSNs against DoS assaults.

Many methods based on machine learning (ML) techniques have been used in the literature in recent years to identify DoS assaults. Much of the research work is used in machine learning (ML) for feature selection methods to get the most performance out of the classifier systems. As a result, there are several ML problems when employing these methodologies. First, there is an enormous quantity of information available on the Internet, which is causing a rapid increase in network traffic. Due to their limited capacity to learn from the features, traditional ML techniques like Support Vector Machines (SVM) are not ideal for the categorization of big data sets. Classical ML produces better outcomes when determining correlations in known assaults than when learning the actions of outliers in unknown harmful attacks. Increases in the number of data transmission networks will increase false alarm rates [12]. DoS attacks seek to disrupt network functionality and deplete the energy and resources of deployed nodes. We created a multi-tier layer architecture based on fog computing. The fog nodes serve as the group's leader, receiving sensed data supplied from the sensor nodes and temporarily storing and analyzing the data.

The proposed framework for this study is made up of three hierarchical layers: Layer 1 is made up of groups of sensor nodes that are randomly placed across the network. They are deployed in the network area to sense the environment and transmit the sensed data to the intermediate layer which is layer 2. Layer 2 consists of a finite number of fog nodes scattered randomly toward the edge of the WSNs. They perform a complementary fusion of all data received from the sensor nodes using event processing techniques, thus allowing local decision-making regarding the sensed area with low latency.

Layer 3 is the topmost layer of the design, which includes data centers for data storage, as depicted in Fig. 1. The data centers are positioned in the cloud to receive data from the fog nodes and store it for further processing [13].

The following are the primary contributions of the proposed scheme:

- The proposed scheme is designed to prevent Selective Forwarding, Black hole, Flooding, and Gray hole attacks in WSNs.
- Flooding and Gray hole attack detection are implemented to prevent channel-based identification within the specified period using the WSNs dataset.
- Decision tree and XGBoost classifier algorithms are used to identify the attacks using the behavior of the sensor node in WSNs.

The remainder of this article is structured as follows:

Section 2 discusses related work; machine learning classifiers are presented in Section 3. Section 4 discusses the Denial-of-Service Attacks (DoS) in WSNs, while the performance evaluation is discussed in Section 5, and the conclusion and future work are discussed in Section 6.

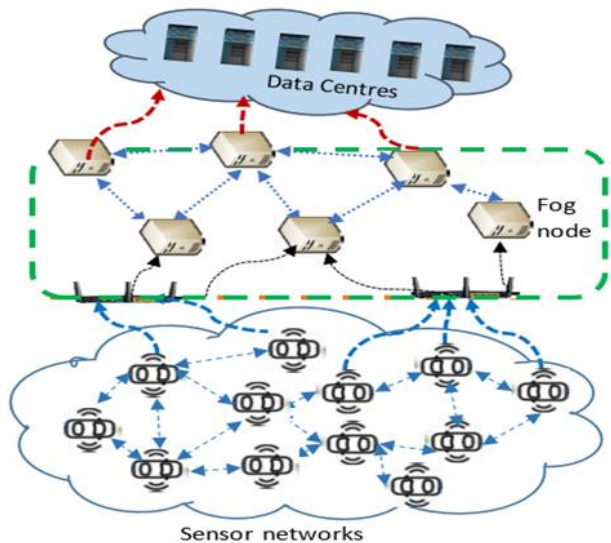

**Fig.1** Framework for the proposed scheme

## 2 Related Work

WSNs are susceptible to various attacks denial of service (DoS) attacks [14–16] due to the mode of data transmission which is through a wireless medium. DoS attacks are major attacks in WSNs, and their main aim is to paralyze the communication network between sender nodes and receiver nodes in the network. Researchers have proposed different approaches in the literature for the detection of DoS attacks in WSNs. In this section, some of the related work in machine learning and Deep Learning is reviewed.

Gunasekarana and Periakaruppanb [17] presented a hybrid protection approach for DoS attacks in WSNs. The authors combined the isolation and routing tables to detect the attack in the network. In addition, a swarm-based defense approach is designed to migrate the faulted channel to a normal operating channel through frequency hop approaches. The proposed scheme is based on a swarm-improved approach for cluster formation and cluster head selection. The results of the study show that the approach saves energy and has low false-positive rates. However, since the approach is based on LEACH, high overhead in the network during the selection of cluster heads is possible.

Li et al. [18] proposed a signal-to-interference-plus-noise ratio (SINR)-based DoS attacks using a game-theoretic approach. The authors first studied an associated two-player game when multiple power levels are available. Thereafter, a Markov game framework was designed to model the interactive decision-making process based on the current state and information collected from previous time steps. They applied a modified Nash Q-learning algorithm to obtain the optimal solutions. The experimental results show that the proposed scheme can detect DoS attacks in WSNs.

Al-issa et al. [19] adopted machine learning techniques (MLDoS) to detect DoS attacks in WSNs. They applied both the Decision Trees and Support Vector Machines (SVM) algorithms to detect attack signatures on a specialized dataset that contains several DoS attacks in WSNs. First, the LEACH protocol is used to collect a dataset containing WSN features in various attacking scenarios. Thereafter, the SVM and decision tree algorithms are used to classify the datasets into normal nodes and compromised nodes. The experimental results indicate that decision trees have higher true positive rates and lower false positive rates than Support Vector Machines, 99.86% versus 99.62%, and 0.05% versus 0.09%, respectively.

Nishanth and Mujeeb presented a detection of flooding attacks in WSNs using the Bayesian Inference method [20]. The proposed method consists of three parts. First, a mathematical model is used to synchronize (SYN) the traffic in the network using Bayesian inference. Second, they developed an efficient algorithm for the detection of SYN flooding attacks using the same approach. Finally, proof of the equivalence of Bayesian inference with exponential weighted moving average was carried out. The results of the proposed method show that it can successfully defend any type of flooding-based DoS attack in WSNs and achieve 98.53% accuracy.

Deep learning based on defense mechanisms for DoS attacks in wireless sensor networks (DLDM) is presented in [21]. The DLDM architecture is based on clustering using homogenous networks. The network is divided into groups depending on the characteristics of the sensor nodes and each sensor node belongs to a cluster that is nearest to it. The proposed approach was implemented and obtained several results through simulation. The results show that the approach can achieve a high detection rate, throughput, packet delivery ratio, and accuracy. This also reduces energy consumption and false alarms in WSNs. However, the number of sensor nodes that were used for the experiment is small. There is a high possibility that the approach may not be effective if the number of nodes is double or three times the number used for the experiment.

Almaslukh [22] proposed a multi-player deep learning and entity embedding-based intrusion detection model for WSNs (DLEID). The approach enables the researcher to effectively learn nonlinear decision boundaries for more accurate classification. Entity embedding automatically learns a robust representation of the raw features with the aid of the deep model. The results of the experiments show that the proposed scheme can transform the raw features into a more robust representation that enables more precise detection of DoS. The summary of the contributions and shortcomings of existing work is presented in Table 1.

**Table 1.** Summary of the contributions and shortcomings of existing work

| Article | Approach | Year | Contribution | Shortcoming |
|---------|----------|------|--------------|-------------|
| [17] | Swarm-based defense | 2017 | The approach can effectively predict the anomalies in a particular cluster | The approach is based on the LEACH protocol, there is a high possibility of high overhead in the network during the selection of cluster heads. The proposed approach is compared to only traditional IDS |
| [18] | Markov Chain Model | 2017 | The approach can reduce energy consumption in sensor networks and minimize false alarms. | No comparison between the proposed and other related work. |
| [19] | Machine learning techniques | 2019 | The authors applied an Intrusion Detection System (IDS) to detect both known and unknown attacks | The approach is based on two weak learnings: decision trees and support vector machines. The experimental results may have low prediction accuracy. |
| [20] | Bayesian inference | 2021 | The authors developed an efficient algorithm for the detection of SYN flooding attacks. | No comparison between the proposed approach and other related work. |
| [21] | Deep learning | 2020 | The approach can accurately isolate the adversaries and it is more resilient to DoS attacks | The approach is compared to only one related work. |
| [22] | data-driven approach | 2021 | The approach can detect a zero-day attack | The authors used raw features to build an intrusion detection model, which can result in low detection accuracy. |
| Proposed | Meta-machine learning | 2023 | The proposed approach can effectively detect all DoS attacks. | |

### 2.1 Assumptions

The following network assumptions are made during the design and development of the proposed scheme.
- Sensor nodes are random and remain static once the nodes are deployed in the network area.
- If a sensor node is compromised, its sensed data will be compromised including the key materials.
- Fog nodes are physical devices with sufficient power and fault-tolerant functionalities.
- Compromised nodes flood bogus route reactions so they are chosen as data forwarders and drop all or some of the data packets.
- All observations in the dataset are dispersed so that the data are not biased.

## 3 Machine Learning Classifiers

This section discusses the use of machine learning models to detect DoS attacks in WSNs. Four machine learning models are presented for the detection of DoS attacks in WSNs. The advantages and disadvantages of each model are described.

### 3.1 Support Vector Machines

A Support Vector Machine (SVM) is a supervised machine learning algorithm for the classification or regression of data groups. The supervised learning systems provide both input and desired output data, which are labeled for classification. SVM works on the principle of fitting a boundary to a region of points that belong to the same class [23]. SVM aims to determine a hyperplane in N-dimensional space (N- the number of features) that uniquely categorizes the data points. The hyperplanes are decision boundaries that assist in classifying the data points. Each data element in SVM denotes a point in the dimensional space with the value of each feature being the value of a particular coordinate. We formulate the SVM optimization problem mathematically as follows.

Minimize
$$\frac{1}{2}\sum_{i=1}^{n} w_i^2 \quad (1)$$

Subject to
$$y_i(\vec{w}\cdot\vec{x_i} + b) - 1 \geq 0 \quad \text{for} \quad i = 1,\dots,N \quad (2)$$

where $w$ is a vector of constants that represent the slopes of the plan; it can be represented with two constants [-1 and 0], $x$ is the data point, $y$ is its corresponding labels, and $N$ is the number of features in the dataset. The main task of the SVM model is to determine the best hyper-plan to classify the data points.

*Advantages of SVM*
*Robustness to noise*: SVM classification algorithms have built-in protection against noise, as the decision boundary is determined by the support vectors, which are the closest data points to the boundary. This approach can provide a unique solution in contrast to Neural Networks where we get multiple solutions corresponding to each local minima for a different sample. Finally, SVM models perform well in high-dimensional space and have excellent accuracy. The model requires less memory because it only uses a portion of the training data.

*Disadvantages*
*Long training time*: Training of the SVM model involves solving the associated dual problem, the number of variables is equal to the number of training data. Thus, if the number of training data is large, solving the dual problem becomes difficult because large storage space will be required, and computation time will increase considerably.

In addition, SVMs have difficulty fitting models on large training sets because solving quadratic programming problems involves computing a matrix of order N. Finally, a more comprehensive discussion of SVMs is required to make them more effective in solving multi-classification problems [24].

### 3.2 Decision Trees

A decision tree is a non-parametric supervised learning algorithm, which utilizes classification for making decisions. It resembles a tree structure to denote the features and outcomes. The decision trees categorize each instance by starting at the root of the tree and progressing until it gets to a terminal node (leaf node) [25]. The root node is further divided into a series of decision nodes (branches) where results and observations are conditionally based. The process of dissecting a node into several nodes is known as splitting. Thus, if a single node is unable to split further into small nodes, then it is called a terminal node. To classify a particular dataset, the decision tree algorithms commence at the root node and progress downward until it gets to a terminal node where a decision is made. The decision tree consists of internal nodes, leaf nodes, and links. A tree node denotes a class of label (output), an internal node denotes the feature (predictor) and the link from a parent node to a child node denotes a decision (rule). Uncertainties could be analytically minimized in a dataset and increase the trustworthiness of the dataset outputs through classification. Algorithm 1 contains the Decision tree algorithm. In addition, this work employs an entropy model to minimize disorder in our target variable. It is mathematically expressed as follows.

$$Entropy = -\sum_{k=1}^{N} p_k log_2(p_k) \quad (3)$$

Entropy denotes the degree of the randomness of category $p_k$.

The uncertainty (impurity) is denoted as the log to base 2 of the probability of a category $(p_k)$, and $p_k$ is the probability of randomly selecting a sample in class $k$. Finally, the decision tree models do not require normalization of data to be performed as they are not sensitive to the variance in the data.

*Advantages of Decision Trees*
A decision tree requires less effort for data preparation during pre-processing. It is a versatile tool that can handle both numerical and categorical data, which can make it useful to solve diverse types of problems.

Secondly, it can be used for regression and classification problems, which means that it can be applied to diverse types of problems.

*Disadvantages of Decision Trees*
Decision trees can be sensitive to minor changes in the data set (instability). A minor change in the data can result in a substantial change in the structure of the decision tree which can produce a different result from what users will get in a normal event.

Secondly, in large datasets, decision trees may not be practical as the number of tree structures grows exponentially with the number of features.

## 3.3 Logistic Regression

Logistic regression is a machine learning algorithm that is used for classification problems. It is a predictive analysis algorithm that is based on the concept of probability [26]. The model is used to study the effects of one or more independent variables on categorical outcomes (dependent variables). It has been used in many sectors including financial institutions to predict credit scoring, medicine to predict the presence and absence of tumors in patients, text editing, gaming, and so on. A statistical model is usually used to compute the binary dependent variable with the help of a logistic function called a sigmoid function. The sigmoid is a mathematical function that accepts a real number and maps it to a probability between 0 and 1. The logistic regression model is expressed as follows.

$$logit = \beta_0 + \beta_1 x_1 + \beta_2 x_2 + \cdots + \beta_N x_N \quad (4)$$

where $x_j$ denotes the predictor (independent) variable, $\beta_0$ denotes the intercept, $\beta_j$ denotes the coefficient estimate for the $j$ predictor variable.

The model computes the probability of an occurrence by passing the outcome of a linear function of features through a logistic function. It then maps the probability to binary outcomes. The sigmoid function is expressed as follows.

$$g(Z) = \frac{1}{1 + e^{-Z}} \quad (5)$$

Thus, if the value of $g(Z)$ goes to positive infinity, then the predicted value of $y$ becomes 1 and if the value goes to negative infinity, then the predicted value of $y$ becomes 0. The simplest idea here is clipping results between 0 and 1. Putting a threshold at 0.5, if the output of the sigmoid function is greater than > 0.5, then the predicted class is "1"; otherwise, the predicted class is "0".

*Advantages of Logistic Regression*

Logistic regression is quite easy to implement and interpret. It provides great training efficiency and less computational power. Moreover, unlike decision trees or support vector machines, this algorithm allows models to be updated easily to reflect new data. The update can be done using stochastic gradient descent.

In addition, it is one of the most effective algorithms when there are linearly separable outcomes or distinctions represented by the data. This means that we can draw a straight line separating the results of a logistic regression calculation.

*Disadvantages of Logistic Regression*

Non-linear problems cannot be solved with logistic regression since it has a linear decision surface. Linearly separable data is rarely found in real-world scenarios. As a result, it is necessary to transform nonlinear features, which can be accomplished by increasing the number of features to linearly separate the data at higher dimensions. Secondly, this algorithm suffers from overfitting for high dimensional datasets resulting in poor generalization to new data.

## 3.4 Extreme Gradient Boosting Algorithm

Extreme Gradient Boosting (XGBoost) is an optimized distributed, large-scale, general-purpose gradient boosting library designed to be highly flexible portable, and efficient [27]. It implements machine learning algorithms under the Gradient Boosting framework to make accurate predictions by combining an ensemble of estimates from a set of weaker and simpler models. The main idea of Boosting is to assign each sample the same initial weight and continually adjust the weights in subsequent iterations. The algorithm is robust and can handle a wide range of distributions, data types, relationships, and hyper-parameters that you can fine-tune. In the gradient-boosting framework, some generalized linear machine learning algorithms and Gradient-Boosted Decision Trees (GBDT) are implemented.

The objective function and the prediction function are usually formulated in supervised learning. The parameters for the training are applied to minimize the objective function to learn the important parameters such that the obtained parameters and the prediction function are used to classify a dataset or numerically predict an unfamiliar sample dataset. The XGBoost model can be expressed as follows.

$$\hat{y}_j = \sum_{k=1}^{N} f_k(x_j) \quad \text{such that} \quad f_k \in F$$
$$\text{for all} \quad j = 1,2,3, \ldots, N \quad (6)$$

where $m$ denotes the number of features, $F = \{f(x) = \omega_g(x)\}$ such that $(g: \mathcal{R}^m \rightarrow \{1,2,3, \ldots, U\}, \omega \in \mathcal{R}^U)$ is the Classification and Regression Trees (CART) structure, $g$ denotes the tree structure of the sample data associated with the terminal nodes, $\omega$ represents the actual score of the real nodes, and $U$ represents the number of terminal nodes.

It is essential to determine the optimal parameters by minimizing the objective function to produce an optimal model. The XGBoost objective function when predicting numerical values can be divided into an error function ($\mathcal{C}$) and a model complexity function ($\xi$). The objective function is expressed as follows.

$$Obj_{fn} = \mathcal{C} + \xi \quad (7)$$

$$\mathcal{C} = \sum_{j=1}^{N} (y_j - \hat{y}_j)^2 \quad (8)$$

$$\xi = \frac{1}{2} \Psi \sum_{v=1}^{U} \omega_v^2 + \eta U \quad (9)$$

where $\eta U$ and $\frac{1}{2} \Psi \sum_{v=1}^{U} \omega_v^2$ denote the regular terms of $\mathcal{C}_1$ and $\mathcal{C}_2$ respectively and $\Psi$ $\eta$ are parameters that prevent the model from overfitting. Thus, when a training dataset is used to optimize the model, it is important to keep the original model unchanged and include a new function $f$ to the model to minimize the objective function. The objective function is formulated as follows.

$$Obj^\delta + = \sum_{j=1}^{N}(y_j - (\hat{y}_j^{(\delta-1)} + f_\delta(\wp_j)))^2 + \xi \quad (10)$$

where the predicted value of the $(\delta - 1)$th model is $\hat{y}_j^{(\delta-1)}$, $f_\delta(\wp_j)$ is the function $f$ added at $\delta$th time and $Obj$ denote the scoring function that can be applied as an evaluation model.

Many regression tree structures can be derived by recursively calling the above tree creation models. $Obj$ is used to search for the optimal tree structure and put it into the existing model to establish an optimal XGBoost model.

*Advantages of XGBoost*

Handling Missing Values: XGBoost is quite easy to implement coupled with the ability to work with a wide range of classifiers and has an in-built capability to handle missing values. If XGBoost encounters a missing value at a node, it attempts both left-hand and right-hand splits and learns the route leading to the best possible loss.

Scalability is another advantage of XGBoost. It is designed for efficient and scalable training of machine learning models, making it suitable for large datasets. It can produce high-quality results in several machine-learning tasks

*Disadvantages of XGBoost*

*Computational Complexity*: A major disadvantage of XGBoost is its computational complexity, especially when training large datasets. As a result, it may not be suitable for systems with limited memory resources.

Lack of transparency is another disadvantage of the XGBoost model: XGBoost has often been considered a "black box" algorithm, which means that it can be difficult to interpret and understand how it arrives at its predictions. This can make it challenging to troubleshoot and fine-tune.

## 3.5 System Model

The architectural framework for the proposed scheme is composed of a randomly distributed set of sensor nodes (sensor networks), fog nodes, and data centers located in the cloud. The sensor networks are modeled as a graph $G = (N, E)$, where $N$ denotes a set of nodes and $E$ denotes the links. The elements of the $N$ are expressed as $N = \{n_1, n_2, n_3, \ldots, n_l\}$ where $l$ is the number of sensor nodes in the network area and $E = \{e_1, e_2, e_3, \ldots, e_l\}$ connotes the communication links between the nodes. An edge joining two nodes $n_i$ and $n_j$ is denoted by $e_{ij}$. We denote $N_i$ to be the set of neighboring nodes of node $n_i$ while $L \in R^{l \times l}$ denotes the Laplacian matrix related to $N$. The initial deployment of WSNs consists of normal nodes and relay nodes and each node has an identification $(ID)$ that uniquely differentiates it from other nodes.

*Relay Nodes:* Energy consumption in WSNs can be conserved during network operation with the inclusion of relay nodes within the network, instead of each sensor node transmitting through a long hop to another node. Thus, relays may be strategically placed among the sensor nodes to break the single long hop into shorter hops. The relay nodes enable sensor nodes to transmit through a short distance. The main aim for using the relay nodes in this work is to minimize overheard that is usually involved during cluster formation and selection of cluster heads in previous studies. It is assumed that each sensor node has a communication range $r_n$ such that $r_n > 0$ and the relay nodes have a communication range $R_n$ such that $R_n \geq r_n$.

Thus, an optimal number of relay nodes needs to be deployed in WSNs so that between every pair of sensor nodes, a communication path consists of sensor nodes and/or relay nodes such that every hop of the path is no longer than the communication range of the sensor nodes and the relay nodes. Due to the mode of deployment of sensor nodes, some nodes could be compromised during network operation and become malicious nodes.

*Normal sensor nodes:* Normal sensor nodes are small, low-powered devices and have a modest processing capability. Their main function is to sense environmental data in their area of deployment and transmit them to the nearest neighboring node or fog depending on the distance between the sender node and the receiver node.

*Normal behavior*: A sensor node behaves normally when it can perform all its functions while maintaining security principles such as integrity, confidentiality, non-repudiation, and authenticity [29].

*Fog computing* is a layered model for allowing ubiquitous access to a shared scalable network of computing resources. The model consists of fog nodes that enable the deployment of distributed latency-aware applications and services. Fog nodes are physical or virtual devices placed between the edge of smart devices and centralized (cloud) services. They are context-aware and support a common communication system and data management. Due to the short transmission range of sensor nodes, fog nodes are deployed close to the edge of WSNs to receive sensed data transmitted from the sensor nodes [30]. The fog nodes can receive and process several types of heterogeneous data generated from diverse types of sensor nodes. In addition, they have more processing power and storage facilities than typical sensor nodes. They are used to connect the sensor networks to the cloud to perform complex computations and generate useful results for end-users.

*Cloud computing*: Cloud computing is the on-demand availability of computer system resources especially data storage (data centers) with high performance, an unlimited power source, and storage without direct active management by the user. It enables the delivery of computing services such as servers, storage (data centers), software, databases, networking, intelligence, and analytics over the Internet [31]. Cloud computing is integrated into our design to receive the heterogeneous data transmitted from the fog nodes for further storage, processing, and analysis in which the results can be used to make informed decisions.

*Prevention of attacks:* Prevention means an attack will not succeed due to the applied, authorized mechanisms that an adversary cannot bypass. Protocols and mechanisms based on prevention strategies can inhibit compromising parts of a computer system. For instance, passwords are used to prevent unauthorized users from accessing the system.

In the context of WSNs, sensor nodes can be easily compromised due to their nature of deployment and mode of transmission [32]. In addition, the nodes are resource-constrained, which exacerbates the use of complex algorithms and protocols. However, it might be possible to employ mechanisms that can be used to preclude an adversary from performing attacks in WSNs. This implies that the mechanisms would protect the sensed data stored on the individual nodes from being retrieved by an attacker.

### 3.6 Energy Consumption Model

Each sensor node in a WSN has a radio communication subsystem consisting of transmitter/ receiver electronics, antennae, and amplifier energy proportional to the receiver distance. We adopted the radio energy dissipation model in [33] for the energy consumption in sensor nodes during transmission. The energy consumed by a sensor node to transmit $q$-bit of sensed data is expressed as follows.

$$E_{Tx}(q,d) = \{q(\Phi + \beta_1 d^2), if\, d < d_0 \wedge q(\Phi + \beta_2 d^4) \quad (11)$$

where $\Phi$ is the electronic energy that depends on factors such as the spreading of the signal, modulation, and digital coding. $\beta_1$ and $\beta_2$ are the amount of energy per bit dissipated in the transmitter amplifier and $d$ is the distance between two sensor nodes as shown in Equation (12).

$$d = \sqrt{(x_i - x_j)^2 + (y_i - y_j)^2} \quad (12)$$

The energy consumed by $E_{Rx}(l)$ to receive the $q$-bit message is given by Equation (13).

$$E_{Rx}(q) = q * \Phi \quad (13)$$

We assume that an attacker node has higher energy than a normal sensor node during the network operation. Based on the energy model in Equations (11) and (13), a node with higher initial and residual energy has a higher probability of becoming a relay node in each round. As mentioned above, nodes in our network are of two types: normal nodes and malicious nodes. The initial energy of a normal node is denoted by $E_0$ and the energy of a malicious node is $(1 + a) * E_0$ which is $(1 + a)$ higher than the energy of a normal node.

Therefore, the total energy consumption for a sensor node is the sum consumed for data transmission and reception as shown in Equation (14).

$$E_{Tot} = E_{Tx}(q,d) + E_{Rx}(q) \quad (14)$$

Considering $d_0$ as the threshold transmission distance for all the sensor nodes, node $j$ is a neighbor node of $i$ if:

$$N_i = \{j : d \leq d_0\} \quad (15)$$

### 3.7 Selection of Relay Node based on Residual Energy

If $R_Z$ denotes the number of rounds to be a relay node for the sensor node, $n_i$ then $P_r(i) = 1/R_Z$ denotes the average probability for $n_i$ to be a relay node during $R_Z$ rounds. $P_r^{opt}$ denotes the optimal probability for a normal sensor node. The probability for a malicious (attacker) node $P_r^{mal}$ is expressed as follows.

$$P_r^{mal} = \frac{P_r^{opt}}{(1+a)} \quad (16)$$

If $E_{Tot}(i)$ represents the energy consumption of node $n_i$ and the average energy is denoted by $E_{avg}$ in the round $R_Z$ of the network, then $P_r(i)$ is expressed as shown in Equation (17).

$$P_r(i) = \begin{cases} \dfrac{P_r^{opt} * E_{Tot}(i)}{(1+a) * E_{avg}} & \text{if i is a normal sensor node} \\ \dfrac{P_r^{opt} * E_{Tot}(i)}{E_{avg}} & \text{if i is a malicious node} \end{cases} \quad (17)$$

where the $E_{avg}$ is expressed as follows.

$$E_{avg} = \frac{1}{N} \sum_{i}^{N} E_{Tot}(i) \quad (18)$$

Based on the above expressions, a compromised node with high remaining energy will have a higher likelihood of being a relay node.

## 4 Denial of Service Attacks (DoS) in WSNs

### 4.1 Flooding Attack

A flooding attack is a type of DoS attack in WSNs. The fundamental problem of the flooding attack is that the compromised (flooder) node ($C_n$) floods the full network by constantly forwarding many advertising messages with a high rate of data transmission ($D_{Tr}$) [34]. The flooder (compromised) node sends message requests to other nodes within the network. Once the node establishes a connection with the legitimate nodes, it continues sending them useless messages until their main source of power (batteries) runs out. The compromised node attempts to deceive the legitimate nodes to choose it as a relay node, particularly those sensor nodes that are within its transmission range to consume their energy and create energy holes among the legitimate nodes.

### 4.2 Black hole attack

A black hole attack is one of the major threats in WSNs. This type of attack occurs during the route discovery phase. A sensor node that is affected by this attack uses its routing protocol to advertise itself as having the shortest path to the data center (destination) and makes all neighboring sensor nodes transmit the data toward itself. In WSNs, sensor data is transmitted from a source node to a destination node. The source node determines its shortest route by forwarding the route request message to all neighboring nodes and each of the neighboring nodes will in turn reply to the sender node with the path replay which consists of its hop count and sequence number. Consequently, if one of the nodes has been compromised by the black-hole attack, the malicious node will send a reply message with the lowest hop number and highest sequence number; therefore, a malicious and fake route will be created. Each of the neighboring nodes will also send its hop counts and sequence number to the sender. Based on the information received by the source node from the neighboring nodes, the source node selects the optimal route by comparing the hop counts and the sequence number of each of these nodes. A sensor node that has been compromised by a black-hole attack has the lowest hop counts and highest sequence number and this information will enable the source node to choose the node and transmit its data through it. Once the malicious node has received the data from the source node, it can drop some or all the sensory data [35].

### 4.3 Selective Forwarding Attack

In a selective forwarding attack, a malicious node behaves like a black hole. The malicious node receives sensed data from member nodes and may drop some or all the data packets, ensuring that the packets are not propagated further. However, such an adversary node runs the risk that neighboring nodes will conclude that it has failed and decide to transmit through other paths. A subtler version of this kind of attack is when a malicious node selectively transmits the data packets. The malicious node can drop or modify the data packets it receives from the legitimate nodes and forward the compromised packets to limit suspicion of wrongdoing. The malicious node prohibits the flow of data packets from legitimate nodes to fog nodes [36].

### 4.4 Gray hole attack

In a Gray hole attack, a malicious node exploits the mode of communication of sensor nodes in WSNs to advertise itself as having a high transmission power, to intercept normal data packets that are passing through it [37]. The attacker nodes may displace their malicious behavior in many ways. In one variation, the attacker node can drop data packets coming from nodes in the network while transmitting all the packets to other sensor nodes. The malicious node exhibits an unpredictable behavior in the way it randomly drops some packets while transmitting other packets, thus making it difficult to detect them. Therefore, there is a need to design efficient algorithms and models that can detect sensor nodes and data packets that have been compromised and prevent them from further taking part in the transmission process in the networks.

### 4.5 Dataset Description

This research utilizes the DoS attacks dataset which was downloaded from CICIDS2017 [38]. The dataset was created by the Canadian Institute for Cybersecurity. The dataset contains 391,752 records of diverse types of DoS attacks. The dataset consists of two classes: 'benign' which means legitimate (normal) data and 'malignant' which means malicious data. It contains 25 attack events and four different types of attacks namely: flooding attacks, black hole attacks, selective forwarding attacks, and gray hole attacks. Commonly, several DoS attack datasets have many constraints like relevant data, data redundancy, and so on. We used 20 events (80%) for training the models and 5 events (20%) for testing. We ensure that at least one attack type is included in the testing data set.

### 4.6 Training and Testing datasets

#### 4.6.1 The Training set

The data set for this study consists of two parts: (i) a training set containing 80% of the data set and it is used to train the machine learning models to identify desired patterns in the data. (ii) a test set containing 20% of the data set and is used to evaluate the performance or accuracy of the model.

To avoid overfitting of training data in this study, cross-validation is applied during the training of the dataset. Overfitting occurs when a machine learning model makes accurate predictions in the training set but cannot generalize and make accurate predictions on the unseen data. Cross-validation is a statistical approach for assessing and comparing machine learning algorithms by partitioning the training set into two parts. The first part is used to train a model and the second part is to validate our model performance during training. This validation process gives information that assists us tune the model's hyper-parameters and configurations accordingly. The model is trained on the training set and concurrently, the model evaluation is performed on the validation set after each epoch.

**Table 2.** Description of Features

| Feature number | Feature name | Symbol | Description |
|---|---|---|---|
| 1 | Sensor node Identification | ID | The ID uniquely distinguishes the individual sensor nodes in the network. |
| 2 | Fog Node Identification | $F_{ID}$ | The $F_{ID}$ is the number that uniquely identifies individual fog nodes. |
| 3 | Time | t | The simulation time of the sensor node. |
| 4 | Normal nodes | $N_n$ | These are normal sensor nodes whose core responsibilities have not been compromised by malicious nodes or other entities. |
| 5 | Compromised nodes | $C_n$ | These are sensor nodes in which attackers have gained control after network deployment to corrupt their sensed data or to saturate the network's capacity. |
| 6 | Relay Nodes | $R_n$ | A relay node receives sensed data from member nodes and transmits the received data to the next node or fog node. It shortens the transmission distance between the distant nodes and the fog nodes. |
| 7 | Number of Relay Nodes | $R_n^u$ | Represents the fraction of the relay nodes in WSNs. |
| 8 | Links between the Sensor Nodes | E | The communication links between the sensor nodes are represented by E. |
| 9 | Energy Consumption | $E_{Tx}$ | The energy consumption in sensor nodes during transmission. |
| 10 | Number of Rounds | $R_{no}$ | A round is an equal period (seconds) allocated to the sensor nodes for data transmission and reception. |
| 11 | Data Transmission rate | $D_{Tr}$ | The rate at which a sensor node transmits its data either to a relay node or to a fog node. |
| 12 | Identification function | $I_{fn}$ | An identification function determines which sensor nodes are compromised in the network. |
| 13 | Data Transmission | $D_{Tx}$ | t is the number of sensed data transmitted by the sensor nodes to the fog nodes. |
| 14 | Data Reception | $D_{Rx}$ | The number of data packets received by each fog node from the sensor nodes at time t. |
| 15 | Rank | $R_k$ | Sensor node order in TDMA schedule. |
| 16 | Send code | $S_c$ | The sending code of the relay node. |
| 17 | Is relay node | $IS_{Rn}$ | Describes if the sensor node is a normal or a relay node. |
| 18 | Sensor node remaining energy | $E_{Rem}$ | Denotes the remaining energy of a sensor node. |
| 19 | Attack type | $A_y$ | Describes the type of attacks in the network. |
| 20 | Sensor node initial energy | $E_{init}$ | Denotes the initial energy of a sensor node. |

### 4.6.2 Cross-validation

Cross-validation is a method for evaluating ML models by training numerous ML models on the training set. The training set is randomly partitioned into K equal-sized folds. A single fold (part) is retained as the validation set for testing the model [28]. The remaining K-1 folds are used to train the model, apply it to the validation set, and record its predictive performance. In this approach, the training and validation sets must cross over in consecutive rounds such that every data point has a chance of being validated. K-fold cross-validation involves repetitive rounds of training and validation such that within each iteration. A different segment of the data is used for validation, while the remaining K − 1 folds are used for learning. We used the default value of K-fold which is 5 folds for our study. If the value of K is larger than 5, it will increase the runtime of the cross-validation algorithm and the computational cost. In the first iteration, the first fold is used to evaluate the model and the remaining four folds are used to train the model. Similarly, in the second iteration, the 2nd fold is used to evaluate the model while the remaining folds are used to train the model. The same process is used for the remaining three folds. The stored predictive performances are then averaged. The optimal model parameter is calculated as the one with the best average prediction performance. Although the model sees the validation set but does not learn from it. The algorithm for the K-fold cross-validation is shown in Table 3.

**Table 3.** Algorithm for the K-fold cross-validation

| S/N | Algorithm for the K-fold cross-validation |
|---|---|
| 1. | Pick the number of folds **K** |
| 2. | Split the dataset into a train and test set |
| 3. | Randomly divide the train set into **K** equal parts |
| 4. | Use **K-1** folds as the training set |
| 5. | Use the remaining fold as the validation set |
| 6. | Train the model on the training set |
| 7. | Validate on the test set |
| 8. | Save the result of the validation |
| 9. | Repeat steps 4-8 K times. Each time use the remaining as the test set and validate the model on every fold |
| 10. | Average the results obtained in step 8. |

### 4.6.3 The Test Set

The test set is a separate set of data used to evaluate the model after completing the training. It provides an unbiased final model performance metric in terms of precision, accuracy, and so on. It ensures that the models can predict correctly with the unseen dataset. As mentioned above, 20% of the data set is used as the test set to test the trained data set. The test set mustn't include any data from the training set. Otherwise, it will be difficult to evaluate whether the algorithm has acquired the ability to generalize from the training set or only learned how to memorize it. An algorithm that generalizes well with the training data will be able to effectively perform a task with the unseen data. A learning algorithm that memorizes the training data through an overly complex model can accurately predict the values of the response variable for the training set but will fail at predicting the values for the test set. Memorizing the training set is known as overfitting.

## 4.7 Feature Selection

The feature selection process in machine learning involves selecting notable features in a dataset that will improve the model's performance and interpretability.

The main aim of feature selection techniques in machine learning is to determine the best set of features from the dataset which enables one to design optimized models of the research being undertaken. Feature selection in ML is used to enhance the performance of ML models. In most cases, all the variables in a dataset are not always useful for building a model. Adding redundant variables reduces the model's generalization capability and may also reduce the overall accuracy of a classifier [39]. Thus, the selection of essential features can improve the performance of ML models. In this study, we use feature selection and multi-features to enhance the performance of learning models.

Principal Component Analysis (PCA) and Singular Value Decomposition (SVD) are used to choose the best features from the dataset. The description of features is presented in Table 2.

### 4.7.1 Principal Component Analysis

The PCA is a feature selection technique for reducing the dataset's dimensionality and increasing interpretability while minimizing information loss. The PCA reduces the dimensionality of a dataset while preserving most of the changes in the dataset for faster and more efficient data processing. This method is highly effective, simple to understand, and does not impose any constraints on the choice of parameters. The central idea of PCA is to recognize the relationship in the data. It maps the features of X-dimensional with the label (y) such that labeled: $\{features, label\}: (X, y)$ to optimize certain variance. PCA minimizes the redundancy of data by constructing new smaller features that have vital information about the original data. Let's say we have a dataset of $p \times n$ matrix, X, the meaning of X-dimensional data features can be expressed as follows.

$$S = \frac{1}{n}\sum_{b=1}^{n} p_i \qquad (19)$$

where $n$ denotes the total number of variables and $p_i$ denotes a data feature.

The covariance ($Cov$) matrix using the value of mean (S) is given as follows.

$$Cov = \frac{1}{n}\sum_{b=1}^{n}(p_i - S)((p_i - S)^T) \qquad (20)$$

The feature vector values, Q, and covariance matrix values are evaluated as follows.

$$Cov = Q.\Sigma.Q^T \qquad (21)$$

$$\Sigma = diag((\sigma_1, \sigma_2, \ldots\ldots, \sigma_n)\sigma_1 \geq \sigma_2 \geq \cdots \geq \sigma_n \geq 0) \qquad (22)$$

where $\Sigma$ is the arranged diagonal matrix of $n$ feature values in descending value and $\sigma_i$ is the corresponding covariance matrix feature value and Q is the feature vector values.

We used both feature values and feature vectors to evaluate the cumulative variance contribution rate for the first Y-row element and express it as follows.

$$R = \frac{\sum_{b=1}^{Y} \sigma_b}{\sum_{c=1}^{n} \sigma_c} \qquad (23)$$

where R represents the cumulative variance contribution rate of the first y-row principle elements.

Thus, the Y-row feature vector is obtained by performing dimension reduction as follows.

$$N = Q_y.X \qquad (24)$$

where $Q_y$ denotes the feature matrix and $\mathbb{N}$ represents Y-dimensional data which is derived after mapping with the X-dimensional of the given dataset. The dimensional reduction is obtained by a linear transformation of data X to $\mathbb{N}$.

### 4.7.2 Singular Value Decomposition

SVD is a feature selection method for representing matrices as a series of linear approximations that expose the matrix's underlying meaning. The SVD aims to find the optimal set of factors that best predict the outcome. It has been extensively used for dimension reduction by decomposing a matrix and exposes many useful and interesting characteristics of the original matrix [40]. The SVD of a matrix $\mathcal{B}$ is the factorization of $\mathcal{B}$ into the product of three matrices $\mathcal{B} = \hat{E} D \hat{O}^T$ where the columns of $\hat{E}$ and $\hat{O}$ are orthogonal matrices and the matrix D is diagonal with positive real entries of matrix $\mathcal{B}$.

$$FP_r = \frac{FP}{FP + TN} \tag{30}$$

### 4.7.3 Multi-feature Selection

This sub-section considers chosen features from both PCA and SVD to get a more significant feature set. We selected 20 features, the first 10 were assigned to PCA, and the last 10 to SVD as shown in the equations below.

$$\text{PCA}_{10} = \text{PCA}_{dataset} \tag{25}$$

$$\text{SVD}_{10} = \text{SVD}_{dataset} \tag{26}$$

$$\text{multi\_feature}_{20} = \text{PCA}_{10} \cup \text{SVD}_{10} \tag{27}$$

where $\text{multi\_feature}_{20}$ is the multiple features obtained from the results of PCA and SVD.
The multi-feature method aims to improve the performance of ML models by choosing only those features that have a high contribution to the prediction. A combination of features from both approaches can optimize the models' performance and reduce their computational complexity.

## 5  Performance Evaluation

To minimize the impact of variability on the training set on the experimental results, we applied the cross-validation discussed in subsection 4.6.2 and the multi-feature discussed in subsection 4.7.3. The experiment is repeated 35 times to ensure consistency of the results and is averaged. The average values are used for the plotting of the graphs.
The following metrics are used to evaluate the performance of the machine learning models used in DoS detection. True Positive rate $(TP_r)$, True Negative rate $(TN_r)$, False Positive rate $(FP_r)$, False Negative rate $(FN_r)$, Accuracy $(ACC)$, Precision, and Mean Square Error $(MSE)$.

*True Positive rate:* denotes the number of attack cases correctly classified as attacks. It denotes the rate at which normal node data is correctly classified as normal data. The rate is mathematically expressed as follows.

$$TP_r = \frac{TP}{TP + FN} \tag{28}$$

*True Negative rate:* denotes the number of normal cases classified correctly as normal. It is the rate at which normal node data is correctly classified as normal data. The rate is mathematically expressed as follows.

$$TN_r = \frac{TN}{TN + FP} \tag{29}$$

*False Positive rate:* the number of normal cases that have been incorrectly classified as attacks. It is also known as the false alarm rate, which measures the rate of no-attack cases that were incorrectly labeled by the classifier as normal data. The rate is mathematically expressed as follows.

*False Negative rate:* the number of attack cases that were classified incorrectly as normal. It denotes the rate of attack cases that are correctly classified as malicious data. The rate is expressed as follows.

$$FN_r = \frac{FN}{FN + TP} \tag{31}$$

*Accuracy*: It is expressed as the proportion of instances that are classified correctly (predicted positive cases) against the total number of instances.

$$ACC = \frac{TP + TN}{TP + TN + FP + FN} \tag{32}$$

*Precision:* Precision is used to measure the accuracy of a model. It is the ratio of true positives (correctly classified) to the total number of positive predictions.

$$Precision = \frac{TP}{TP + FP} \tag{33}$$

*Recall:* It is the ratio between all the true positives and the sum of the true positives and false negatives. They are respectively expressed as follows.

$$Recall = \frac{TP}{TP + FN} \tag{34}$$

*F1-score:* It is a measure of a test's accuracy. It is calculated as the harmonic mean of the precision and recall scores. It is mathematically expressed as follows.

$$F1 - Score = \frac{TP}{TP + \frac{1}{2}(FP + FN)} \tag{35}$$

*Mean Squared Error (MSE)* is computed as the mean of the squared differences between predicted and expected target values in a dataset. An MSE value of zero indicates there was no variance between the predicted and expected values. Thus, values close to zero indicate a small variance between the original output values and the expected output values for every record in the training or test set. It is mathematically expressed as follows:

$$MSE = \frac{\sum_{k=1}^{S}(E_k - T_k)^2}{S} \quad (36)$$

where $E_k$ is the expected values, $T_k$ represents the predicted values of a variable, and S is the number of observations available for analysis. $TP$ is the number of malicious sensed data correctly classified as malicious data. $TP$ is the number of normal datasets wrongly classified as malicious data. We determine the cost function and compare it with other related approaches for performance evaluation. The cost function is evaluated based on several misclassified datasets and it is mathematically expressed as follows.

$$Cost = 1 - \frac{TP}{TP + FN} + \emptyset\left(\frac{FP}{FP + TN}\right) \quad (37)$$

where $\emptyset$ denotes the cost difference between the misclassified dataset and the false alarm. Thus, the lower the value of the cost function, the better the performance of detection.

*Confusion Matrix:* A confusion matrix presents a table layout that is used to define the performance of a classification algorithm and helps visualize its outcomes. When the number of classes (labels) in a dataset is more than one, using the classification accuracy alone can be misleading. Thus, using a confusion matrix can be a better indicator of knowing the number of outcomes that were predicted correctly and the number of outcomes that were wrongly predicted.

**The receiver operating characteristic (ROC)** curve is a graph that is used to analyze the performance of a classification model at all classification thresholds. It uses a true positive rate and a false positive rate to characterize the classifier's performance. It shows the true positive rate (sensitivity) at the y-axis plotted against the false positive rate (x-axis) at various cut-off points of the assessment.

## 5.1 Discussion of the experimental results

The performances of the detection rate for the four models (SVM, Logistic Regression, Decision Tree, and XGBoost) discussed above are presented in this section and the performance of each algorithm is presented in Table 4. The experimental results show that the XBoost model performed best using all the metrics with a 98.4% accuracy. The Decision Tree model provided an accuracy rate of 95.8% as the second highest performing model followed by the Logistic Regression model with an accuracy of 91.3%. Among the four models, the SVM model had the lowest performance with an accuracy of 82.4%. Similarly, the XGBoost model has the highest percentage of F1-Score in each of the attacks with Normal (99.5%), Selective Forwarding (96.6%), Black hole attack (96.0%), Flooding attack (95.7%), and Gray-hole (95.6%).

**Table 4**. Analyses of the performance of individual algorithms on training data

| Algorithm | Attack type | Recall % | Precision % | F1-Score % | Accuracy (%) |
|---|---|---|---|---|---|
| SVM | Normal | 99.3 | 99.0 | 99.1 | |
| | Selective Forwarding | 76.5 | 83.4 | 79.8 | |
| | Black hole | 82.4 | 85.9 | 84.1 | |
| | Flooding | 79.6 | 82.1 | 80.3 | |
| | Gray hole | 86.1 | 87.0 | 86.5 | 82.4 |
| | Mean | 84.8 | 87.5 | 86.1 | |
| Logistic regression | Normal | 99.5 | 89.7 | 94.3 | |
| | Selective Forwarding | 96.7 | 97.5 | 97.1 | |
| | Black hole | 92.4 | 91.2 | 91.8 | |
| | Flooding | 89.0 | 93.7 | 91.6 | |
| | Gray hole | 91.9 | 89.1 | 90.5 | 91.3 |
| | Mean | 93.9 | 92.2 | 93.1 | |
| Decision Tree | Normal | 99.5 | 99.1 | 99.3 | |
| | Selective Forwarding | 97.0 | 96.8 | 96.9 | |
| | Black hole | 95.6 | 94.3 | 94.9 | |
| | Flooding | 92.1 | 96.0 | 94.0 | |
| | Gray hole | 94.2 | 93.7 | 93.9 | 95.8 |
| | Mean | 95.7 | 96.0 | 95.8 | |
| XGBoost | Normal | 99.7 | 99.3 | 99.5 | |

|  | Selective Forwarding | 96.8 | 96.4 | 96.6 |  |
|---|---|---|---|---|---|
|  | Black hole | 97.1 | 95.0 | 96.0 |  |
|  | Flooding | 94.3 | 97.1 | 95.7 |  |
|  | Gray hole | 96.0 | 95.3 | 95.6 | 98.4 |
|  | *Mean* | 96.8 | 96.6 | 96.7 |  |

We performed more experiments to further evaluate the models on testing set data and the results are presented in Table 5. The XGBoost model performed best with an accuracy score of 98.5%, followed by the Decision Tree model with an accuracy of 97.3%, Logistic Regression with 91.8%, and SVM with 81.7%. The performance of the models is like the results we obtained from the training data above.

The XGBoost model has the highest percentage of F1-Score in each of the attacks with Normal (99.5%), Selective Forwarding (97.9%), Black hole attack (98.7%), Flooding attack (97.9%), and Gray-hole (97.2%).

**Table 5.** Analyses of the performance of individual algorithms on testing data

| Algorithm | Attack type | Recall % | Precision % | F1-Score % | Accuracy (%) |
|---|---|---|---|---|---|
| SVM | Normal | 99.4 | 99.0 | 99.2 |  |
|  | Selective Forwarding | 76.9 | 83.8 | 80.2 |  |
|  | Black hole | 81.7 | 83.1 | 82.4 |  |
|  | Flooding | 80.4 | 85.6 | 82.9 |  |
|  | Gray hole | 82.9 | 86.3 | 84.6 | 81.7 |
|  | *Mean* | 84.3 | 87.6 | 85.9 |  |
| Logistic regression | Normal | 99.7 | 89.7 | 94.4 |  |
|  | Selective Forwarding | 87.2 | 86.5 | 86.8 |  |
|  | Black hole | 92.9 | 91.2 | 92.0 |  |
|  | Flooding | 89.3 | 93.7 | 91.4 |  |
|  | Gray hole | 91.4 | 89.1 | 90.2 | 91.8 |
|  | *Mean* | 92.1 | 92.2 | 91.1 |  |
| Decision Tree | Normal | 98.8 | 90.0 | 94.2 |  |
|  | Selective Forwarding | 97.2 | 96.3 | 96.7 |  |
|  | Black hole | 98.9 | 96.2 | 97.5 |  |
|  | Flooding | 97.1 | 96.8 | 969 |  |
|  | Gray hole | 98.6 | 97.5 | 98.0 | 97.3 |
|  | *Mean* | 98.1 | 95.4 | 96.7 |  |
| XGBoost | Normal | 99.8 | 99.3 | 99.5 |  |
|  | Selective Forwarding | 97.6 | 98.3 | 97.9 |  |
|  | Black hole | 99.3 | 98.2 | 98.7 |  |
|  | Flooding | 98.7 | 97.1 | 97.9 |  |
|  | Gray hole | 96.9 | 97.5 | 97.2 | 98.5 |
|  | *Mean* | 98.5 | 98.1 | 98.2 |  |

### 5.2  Decision Tree using a complete testing set

Based on the results obtained above, we selected the two most performing models (Decision Tree and XGBoost) with the highest accuracy to know the number of the dataset that is correctly classified and wrongly classified with the errors involved in the classification. Table 6 presents the summary of the experimental results using the complete testing set to evaluate the accuracy of the proposed scheme using a decision tree. The decision tree model classified 73265 datasets correctly achieving 97.3% and misclassified 2115 datasets (2.7%) which coincides with the result obtained in Table 5.

**Table 6.** Decision Tree results from the summary using the complete testing set

| Dataset classified Correctly | 76235 | 97.3% |
|---|---|---|
| Dataset classified incorrectly | 2115 | 2.7% |
| Mean absolute error | 0.003 | |
| Root mean squared error | 0.2146 | |
| Relative absolute error | 1.8652 | |
| Root relative squared error | 13.6281 | |
| Total number of datasets | 78350 | |

In Table 7, we present the confusion matrix for the Decision Tree model when tested with the complete testing set. The rows denote the output class (predicted class), and the columns denote the target class (target class). The diagonal elements represent the number of observations for which the predicted label is equal to the true label (correctly classified), while off-diagonal elements are misclassified by the decision tree algorithm.

**Table 7.** Decision Tree confusion matrix using the complete dataset

| Category | Normal | Selective Forwarding | Black hole | Flooding | Gray hole |
|---|---|---|---|---|---|
| Normal | 18491 | 92 | 178 | 163 | 105 |
| Selective Forwarding | 98 | 13638 | 103 | 99 | 153 |
| Black hole | 158 | 101 | 15879 | 147 | 173 |
| Flooding | 149 | 156 | 106 | 13383 | 152 |
| Gray hole | 96 | 104 | 158 | 87 | 10381 |

### 5.3 Decision Tree using designated attacks

DoS attacks can affect WSNS to various degrees. Flooding has been considered the biggest DoS attack for WSNs as it is the most treacherous attack since it is extremely hard to uncover [20]. For instance, an adversary can announce a very high-quality route to the data center to individual nodes in the network could cause many nodes to attempt to transmit their sensor data through this route. Those nodes that are sufficiently far away from the adversary node would be transmitting sensor data into oblivion, wasting the energy of the sender nodes, and the network is left in a state of confusion.

A gray hole is another dangerous DoS attack in WSNs. It is an advanced transformation of a black hole attack. An adversary node may appear as a normal node and sometimes, it will be manifested as an adversary node based on the attack and either randomly or constantly drops some sensor data that it receives from the neighboring nodes and thus reduces the efficiency of the WSNs. This inconsistent behavior of gray hole nodes makes it difficult to detect the attack in WSNs.

We performed experiments that included only flooding and gray-hole attacks to test the accuracy of the decision trees. The summary of the results is presented in Table 8. The decision trees model correctly classified the 76391 datasets (97.5%) and misclassified the 1959 dataset (2.5%) and the confusion matrix is presented in Table 9.

**Table 8.** Decision Tree results from the summary using designated attacks

| Dataset classified Correctly | 76391 | 97.5% |
|---|---|---|
| Dataset classified incorrectly | 1959 | 2.5% |
| Mean absolute error | 0.001 | |
| Root mean squared error | 0.1837 | |
| Relative absolute error | 1.2931 | |
| Root relative squared error | 8.04383 | |
| Total number of datasets | 78350 | |

**Table 9.** Decision Tree confusion matrix using designated attacks

| Category | Normal | Flooding | Gray hole |
|---|---|---|---|
| Normal | 32142 | 104 | 75 |
| Flooding | 63 | 25985 | 49 |
| Gray hole | 59 | 71 | 19802 |

### 5.4 Extreme Gradient Boosting (XGBoost) with a complete testing set

The experimental results of the XGBoost model obtained are presented in this section. First, we discuss the XGBoost results with the complete testing set and thereafter we discuss the results obtained with the designated (selected) attacks. The summary of the experimental results is presented in Table 10. In this experiment, the full dataset is used to test the accuracy of the scheme using XGBoost. The confusion matrix with the complete dataset is shown in Table 11.

**Table 10.** XGBoost results summary using the complete testing set

| Dataset classified Correctly | 77018 | 98.3% |
|---|---|---|
| Dataset classified incorrectly | 1332 | 1.7% |
| Mean absolute error | 0.001 | |
| Root mean squared error | 0.183 | |
| Relative absolute error | 1.293 | |
| Root relative squared error | 8.0438 | |
| Total number of datasets | 78350 | |

**Table 11.** XGBoost confusion matrix using the complete testing set

| Category | Normal | Selective Forwarding | Black hole | Flooding | Gray hole |
|---|---|---|---|---|---|
| Normal | 25820 | 168 | 94 | 152 | 95 |
| Selective Forwarding | 65 | 8716 | 53 | 42 | 34 |
| Black hole | 84 | 32 | 17356 | 75 | 29 |
| Flooding | 65 | 48 | 62 | 13564 | 68 |
| Gray hole | 39 | 27 | 39 | 61 | 11562 |

### 5.5 XGBoost with a dataset of designated attacks

We performed experiments that included only flooding and gray-hole attacks on the testing set of the XGBoost model. The summary of the experimental results is presented in Table 12. The number of datasets correctly classified datasets is 77574 (99.01%) and misclassified 776 datasets (0.99%) and the confusion matrix is presented in Table 13.

**Table 12.** XGBoost results summary using the designated attacks

| Dataset classified Correctly | 77574 | 99.01% |
|---|---|---|
| Dataset classified incorrectly | 776 | 0.99% |
| Mean absolute error | 0.001 | |
| Root mean squared error | 0.175 | |
| Relative absolute error | 1.274 | |
| Root relative squared error | 7.253 | |
| Total number of datasets | 78350 | |

**Table 13.** XGBoost confusion matrix using the designated attacks

| Category | Normal | Flooding | Gray hole |
|---|---|---|---|
| Normal | 32672 | 104 | 127 |
| Flooding | 114 | 24853 | 173 |
| Gray hole | 106 | 152 | 20049 |

### 5.6 Average Energy Consumption

We analyze the network lifetimes of the proposed scheme with the related approaches discussed in Section 2 such as DLEID, DLDM, and MLDoS in terms of energy consumption. The dissipation of energy has been constantly observed in the process. The experiment was run over 2500 rounds and the average results were determined to ensure consistency of the results. In all the approaches considered, the average energy consumption increases as the number of rounds increases. The proposed scheme consumes the least amount of energy compared to the DLEID, DLDM, and MLDoS respectively as depicted in Fig 2. This is due to the elimination of clustering and selection of cluster heads which increase overheads in terms of computation and inter-cluster communication cost resulting in energy dissipation during network operation. In addition, the integration of fog nodes into the proposed scheme enables sensor nodes to transmit sensor readings only while processing of the readings and other services are performed by fog nodes. The proposed scheme can conserve energy consumption by 50% compared to the three related approaches.

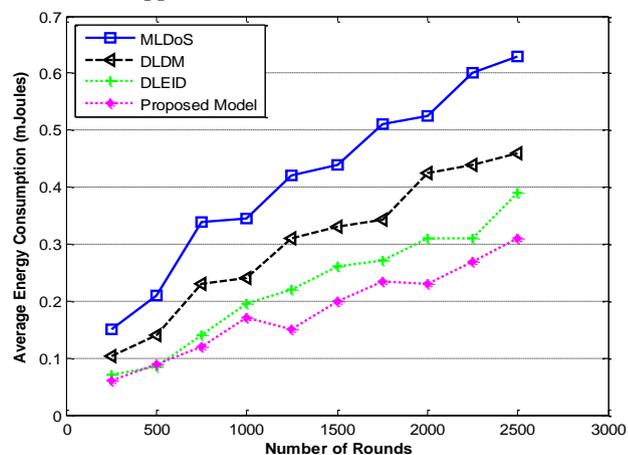

**Fig. 2**. Average Energy Consumption versus Number of Rounds

### 5.7 Network Lifetime

The network lifetime for the proposed scheme and the three other related schemes: DLEID, DLDM, and MLDoS are determined using descriptive statistics and the result is presented in Table 14.

The proposed scheme has a longer network lifetime compared to the three schemes. The reason is that fog nodes are strategically placed in layer 2 of the framework which act as group heads and can receive sensor data from the lower layer. Sensor nodes do not need to run an algorithm to choose a node to be a cluster head among themselves and able to conserve energy. Therefore, extends the network lifetime utilization.

Table 14: Descriptive statistics of the lifetime of the network

| Approach | No. of nodes | Mean | Standard Deviation | Standard Error | 95% Confidence Interval for Mean | | Minimum Lifetime (ms) | Maximum Lifetime (ms) |
|---|---|---|---|---|---|---|---|---|
| | | | | | Upper Bound | Lower Bound | | |
| Proposed Scheme | 200 | 896.13 | 426.21 | 30.14 | 1329.37 | 462.89 | 258.00 | 1451.00 |
| DLEID | 200 | 786.63 | 371.95 | 26.30 | 1167.31 | 405.94 | 250.00 | 1254.00 |
| DLDM | 200 | 650.47 | 295.80 | 20.92 | 925.58 | 375.35 | 250.00 | 1195.00 |
| MLDoS | 200 | 595.58 | 272.67 | 19.20 | 875.32 | 315.84 | 250.00 | 1182.00 |

## 5.8 Throughput

The network throughput can be defined as the amount of sensor data successfully delivered over a specified period from a source node to a destination node. It is measured as kilobits per second (Kbps) or megabits per second (Mbps). It can be expressed as

$$\text{Throughput} = \frac{\mathcal{B}}{\mathcal{T}} * 100 \qquad (38)$$

where $\mathcal{B}$ represents the total number of data delivered to the data center and $\mathcal{T}$ represents the total time taken to deliver the data.

The throughput of the proposed scheme is compared with DLEID, DLDM, and MLDoS with respect to the network size are presented in Fig 3. It can be observed that the throughput increases in each of the approaches as the number of nodes increases. The results show that the proposed scheme has a higher throughput compared to the selected approaches. Thus, the number of sensor data delivered to the fog nodes increased. In the proposed algorithm, many sensor readings were successfully transmitted to the fog networks compared to the selected schemes. The reason is that each sensor node transmits its readings through short distances and each sensor node's role is based solely on the transmission of sensor readings.

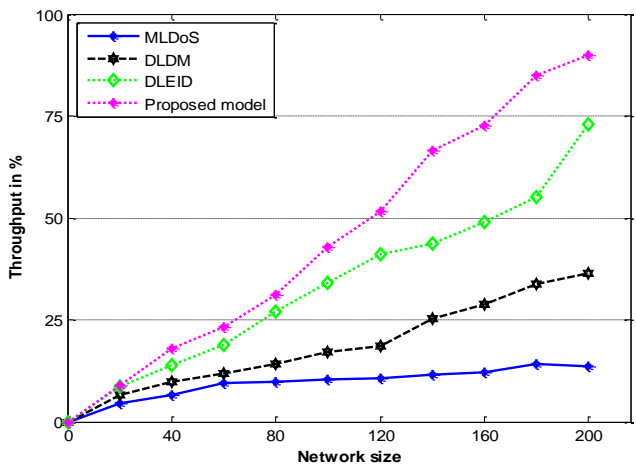

**Fig. 3**. Throughput of the proposed scheme with selected approaches

## 6   Conclusion and Future Work

This research aims to design an efficient WSNs DoS attack detection scheme. To achieve this goal, a specialized WSNs dataset was used for this research. The dataset consists of a normal dataset and four distinct types of DoS attacks namely Selective Forwarding, Black hole, Flooding, and Gray hole. We experimented with four machine learning models namely Support Vector Machine, Logistics Regression, Decision Tree, and XGBoost. Cross-validation was used to avoid overfitting during the machine learning models' training. From the results of the experiments, the XGBoost model achieved the best performance with an accuracy of 98.3%. In the future, we intend to use more datasets for further detection of DoS attacks in heterogeneous wireless sensor networks using other types of DoS attacking scenarios and protocols.

## Declaration of competing interest

The authors confirm there are no competing interests between them and their organizations.

## Author's contributions

A.P. Abidoye conceived the idea and wrote the abstract, the introduction, and the methodology. I.C. Obagbuwa performed the experiments and presented the proposed solutions. N.A. Azeez wrote the related work and the conclusion. All the authors contributed to the draft of the manuscript of the submitted paper.

## Data availability

Data will be made available on request.